# Once FITS, Always FITS? Astronomical Infrastructure in Transition

M. Scroggins, B. Boscoe

**Abstract**— The FITS file format has become the de facto standard for sharing, analyzing, and archiving astronomy data over the last four decades. FITS was adopted by astronomers in the early 1980s to overcome incompatibilities between operating systems. On the back of FITS' success, astronomical data became both backwards compatible and easily shareable. However, new advances in astronomical instrumentation, computational technologies, and analytic techniques have resulted in new data that do not work well within the traditional FITS format. Tensions have arisen between the desire to update the format to meet new analytic challenges and adherence to the original edict for FITS files to be backwards compatible. We examine three inflection points in the governance of FITS: a) initial development and success, b) widespread acceptance and governance by the working group, and c) the challenges to FITS in a new era of increasing data and computational complexity within astronomy.

---

## 1 CAN A FILE FORMAT GOVERN?

*I may be blasphemous here [on the Vatican adopting FITS], I [think], "Oh that's another sign that things are in bad shape".… Maybe we're being a little too conservative here – Astronomer [1]*

In the late 1970s, a handful of astronomers came together to build a file format specifically designed for astronomical data. There is nothing unusual about this in and of itself; scientists commonly built formats, readers, and other low-level tools to do their science. However, the Flexible Image Transport System (FITS) was guided by an unbreakable rule inscribed in the format: it must be backward compatible to allow for the future interchange of astronomical observations. Stability brought by this rule has enabled FITS to be still used today by astronomers world-wide for all types of data. Inspired by this stability, in 2014 the Vatican decided to adopt the format for their image archives [2]. Yet, a stable archival format is not necessarily what all astronomers would like; some want a format that has adapted to modern computing capabilities, of which FITS lacks.

Since 1982, the FITS file format has been the de facto standard for sharing, analyzing, and archiving astronomical data. The format was formally described in a 1981 paper [3], and widely adopted by astronomical observatories and scholarly associations shortly thereafter because of its ability to overcome the non-interoperability of contemporary computer systems. FITS was an immediate success and within a few short years, working astronomers were able to exchange data with colleagues regardless of which computer systems they used or whether their images were generated via radio or optical telescope. By 1990, when NASA's Hubble telescope reached first light and the nascent observations were disseminated via FITS files, FITS had become the lingua franca of astronomy, as unavoidable within the confines of astronomy as English is within the broad expanse of science [4].

Along with exchange came a design maxim and policy edict – once FITS, always FITS. FITS was to remain forever backwards compatible and new versions of FITS were not to break the maxim. The edict is a modern formulation of a tradition borne of the demands and hard experience of observational science; rare events observed at great difficulty are to be kept accessible for future astronomers. FITS has changed rarely, and then only in small increments. Since 1993 only seven modifications have been made to the format [5] and only five official modifications to the standards of the format have been made. In keeping with astronomical tradition, which gives the observer control of the mechanics and tooling of the telescope used for observation, FITS remains open to local modifications and workarounds, hence the term "Flexible" in the name.

Accompanying FITS, and coterminous with its wide adoption, came a specific form of governance. Between 1988 and 2016, FITS was governed by an official International Astronomy Union (IAU) working committee, the FITS Working Group (FWG), before being superseded by the Data Representation Working Group (DRWG) [6, p. 2]". As an international organization, the IAU FWG encompasses numerous national and supranational astronomical organizations, such as the American Astronomical Working Group on Astronomical Software (WGAS) and the European FITS Committee. Meetings are held on an ad hoc basis as comments, objections, queries, and suggested changes to FITS from the polity of working

---

- *M. Scroggins is with the Department of Information Studies, University of California, Los Angeles, CA 90095. E-mail: mscroggins@ucla.edu.*
- *B. Boscoe is with the Department of Information Studies, University of California, Los Angeles, CA 90095. E-mail: boscoe@ucla.edu.*

astronomers filter up to the FWG from conferences such as the Astronomical Data Analysis Software and Systems (ADASS) Birds of a Feather (BoF) sessions, feedback from newsgroups, e-mail exploders and informal conversations between colleagues [7].

Governance entered the social science lexicon in the late 1980s as a way to account for the decoupling of policy from the traditional organs of government [8]. While academic governance has long been a topic of scholarly concern, scholars have more recently turned their attention to the governance of open source software [9]. FITS is unusual in being a publically shared, non-commercial format governed by scholarly associations. To account for FITS' dual parentage, we take up FITS' governance as a form of material politics through the lens of infrastructure. We examine how the governance of FITS is inscribed in technical standards, astronomical institutions, and in the everyday use of FITS, and consider how FITS is made durable by this infrastructure and how, in turn, infrastructure has FITS' place in astronomy durable.

Infrastructure, as Larkin [10, p. 328] argues, constitutes an "architecture for circulation" that fades into the background (*infra-* from the Latin for below, or, out of sight) as it foregrounds an object of circulation. Like a road, that circulates traffic while the road qua road settles into the background, infrastructures remain hidden in ordinary use until a breakdown, a pothole for instance, brings them to our attention [11]. Infrastructures have a life, a biography; a beginning and an end [12]. Infrastructure is, therefore, an example of what STS (Science and Technology Studies) theorists have called material politics [13]. Through material politics, such mundane objects as a speed bump, a "sleeping policeman" [14, p. 186] and computer languages, social relations are made durable and persistent through time and across space.

In this paper, we look at the governance of FITS, accounting for FITS meaning for astronomy today: to some, an aging format in need of updating, to others, a rock of stability in a changing analytical environment. In the four sections which follow we (i) recount the conditions of FITS's invention and diffusion within astronomy, (ii) examine the material politics and infrastructure of FITS governance, (iii) discuss the reappearance of the interchange problem within astronomy, and (iv) conclude with observations on the governance and material politics of emerging forms of astronomical infrastructure.

## 2 Don Wells' vision of an astronomical interchange format

*A unique interchange format needs to be very flexible… It should provide a mechanism for transmitting any auxiliary parameters that are associated with the image, even though not all these parameters, nor even the nature of all these parameters, can be specified a priori. – Wells et al. 1981*

From the beginning of astronomy, detailed metadata chronicling positions, time, and conditions have accompanied observations. The earliest form of astronomical observations were drawings. Beginning in the mid-19th century, optical plates slowly replaced drawing as the observation format of record. In the middle of the $20^{th}$ century advances in detecting radio waves made another form of astronomical imaging possible. Growing out of research dating from the 1930s, radio astronomy measures radio waves emitted by celestial objects. It languished until computers powerful enough to handle the Fourier transform inversions necessary to produce images from radio waves came into common use in the late 1960s. When the first radio telescope sky surveys appeared, comparing them to existing optical surveys required a lengthy and error-prone process of fitting images together through trial and error [15]. By the late 1970s astronomy was faced with two observational paradigms - optical and radio - developing along divergent paths.

Meanwhile, the computational environment of the 1970s was a collage of rival manufacturers and proprietary operating systems. On the cusp of the PC era, minicomputers and mainframes dominated the academic computing landscape: mainframes from IBM, Burroughs & UNIVAC, NCR, Control Data Corporation, Honeywell, Cray Research, Digital Equipment Corporation, Hewlett-Packard, Amdahl Corporation, and International Computers Limited; minicomputers from Digital Equipment Corporation, Wang Laboratories, Data General, Apollo Computer, and Prime Computer [16]. This babel of manufactures and incompatible physical file formats meant that sharing data was onerous, at best.

Yet before the problem of incompatible machines could be faced, the logistical problem of transporting observational data from telescope to home institution had to be overcome. In the late 1970s, it was common for observatory computing systems to be incompatible with university computing systems. After the long trek off the mountain with a physical box containing tapes and floppy disks, astronomers were then faced with the problem of making their data available to themselves for analysis.

Out of this chaotic environment, FITS emerged through a serious of informal conversations between astronomers seeking to circulate data. For FITS, the most important of these conversations occurred in 1976, when Ron Harten, a radio astronomer from the Netherlands visited Donald Wells, an optical astronomer, at Kitts Peak Observatory, and began an ongoing conversation about an interchange standard within astronomy that turned into a working format [17, p. 251]. From their conversation, Wells later derived the three mandates that characterize FITS success as an interchange format: enable radio and optical images to be combined into a single file, be governed through international committees, and remain forever backwards compatible [18].

Several key design decisions were also made at this time, the most controversial involving the file header. The header file was to contain human readable info such as when and with what instrument an image was taken. Wells wanted a flexible file header, rather than fixed fields, that accomplished two things: (i) allow FITS to be used by the widest possible audience of astronomers, and (ii) give knowledgeable programmers enough latitude to implement customized solutions. Because they wanted to reach compromise within the polity of working astronomers, Wells and Harten, joined by the radio astronomer Eric Greisen, prototyped FITS using an IBM 360 for the radio data and a CDC 6400 for the optical data, which at the time were known to be incompatible systems. When interchange between radio images generated on an IBM 360 and optical images generated on a CDC 6400 was successful, the foundation for the digitization of astronomy had been laid.

Though the technical aspects of FITS were largely worked out between Wells and Harten, the problem of generating buy-in and support from the larger polity of working astronomers remained. In short order, through Greisen, FITS gained support from National Radio Astronomy Observatory, and backing from NASA. With observatories quickly recognizing the advantages of adopting FITS and several leading observatories already using FITS, the logical next step was to move governance of FITS from an informal, ad hoc, conversational basis to a structure with formalized decision-making norms and institutional support. In this effort, FITS was helped by astronomy's long history of internationalism.

The logical institution to lead the governance of FITS was the International Astronomy Union (IAU). The IAU was formed in 1919 to further international cooperation and communication in the wake of the Great War [19]. Using the IAU's supra-national position within astronomy and well-established system of hierarchical working groups gave FITS a traditional, durable, and far reaching infrastructure for governance. At the 1982 IAU General Assembly meeting in Patras, Greece, Commission 5 (Documentation and Astronomical Data) adopted the following resolution:

> 11. Resolutions submitted by Commissions or by associated Inter-Union Commissions
> The following resolutions have been proposed:
> 
> 1. Resolution proposed on behalf of Commission 5:
>    Commission 5
>    *considering* the present unsatisfactory situation of the transfer of astronomical data between astronomical institutions
>    *recommends* that all astronomical computer facilities recognise and support the Flexible Image Transport System (FITS) for the interchange of binary data on magnetic tape, as described in Astronomy and Astrophysics Supplement, vol. 44, p.363 and 371.

Fig 1. Proposal to the XVlllth General Assembly of the International Astronomical Union [20, p. 16]

The two papers referenced in the resolution constitute (i) the initial definition of the FITS in 1981 and (ii) the initial extension to the FITS format for handling random groups [21]. The years between 1982 and 1990 saw the steady adoption of FITS for everyday work in astronomy, led by a steady stream of software for working with FITS files and the adoption of FITS as an output format by major observatories [22]. In 1990, NASA formally adopted FITS as the standard format for all NASA-funded astrophysics projects. Also in 1990, the American Astronomical Society (henceforth AAS) WGAS FITS committee was established with Don Wells at its head [23].

In support of this new committee, Wells created a listserv to solicit ideas and encourage discussion within astronomy about FITS. Another important venue for soliciting ideas and discussion about FITS established at this time was the BoF group at the newly formed Astronomical Data Analysis Software and Systems (ADASS) conference. Because of its tight focus on astronomical computation and cross-over with IAU and WGAS members, it quickly became the venue of choice for prototyping modifications and extensions to FITS. With the institutional support afforded by NASA's adoption of FITS, the creation of the listserv in 1990, and the initial ADASS conference in 1991, the infrastructure of FITS governance was in place. FITS would be governed by the hierarchical style of academic governance – conferences, committees, working groups, and international associations – filtering suggestions for modifications and improvement upwards to the supra-national IAU FWG.

Before continuing, we must clarify a few specific technical features of FITS. The workflow imagined by Wells in his 1981 framing paper on FITS was a linguistic process of translation into FITS from institution A's internal format then reverse translation from FITS into institution B's internal format. Being a grammar for interchange and interoperability, FITS attempts to sidestep the thorny problem of semantics, the meaning and interpretation of data contained within the file [24]. But one area where FITS blurs the line between acting strictly as a grammar for interchange and semantics is in the file header.

The FITS header is taken directly from the ANSI FORTRAN 1977 standard for list input and takes the form of: keyword = value/comment. The FITS header is made of card images, each card taking 80 columns (bytes), originating from systems using 80 column punchcards. A keyword is any 8-character ASCII string. The value field conforms to the FORTRAN standard. The comment is a human readable textual annotation on the intention, meaning, and use of the keyword/value pair. The header is where Wells et al. placed the responsibility for explaining parameters that cannot be specified a priori and where the syntactical and semantics elements of FITS become blurred. As Wells et al. observed of FITS headers "coordinate information and auxiliary parameters are important for the unambiguous interpretation of the digital image, particularly when the object of exchange is the intercomparison of sources as seen by various detector systems" [3, p. 365].

The lacunae formed in the FITS header system would turn FITS into a format of dialects and creoles [25]. All can speak to each other but require a great deal of translation. On a syntactic level, FITS has successfully bridged the differences between optical and radio astronomy and tamed the chaos in computing hardware and software. But delegating the responsibility of explaining the semantic elements of a file, a task requiring contributory expertise in astronomy [26], to the header had the effect of placing a limit on the kind of intra-astronomical interchanges FITS could accomplish. In a later section, we will take up how the interpretation of digital images within astrono-my became problematic when the World Coordinate System (henceforth WCS) was introduced into FITS and more recently in the development of complex analytical pipelines within astronomy.

## 3 FITS and the Circulation of Astronomical Observations

*My intention in forming the newsgroup is that _anyone_, _anywhere_ is welcome to discuss FITS in this place, and that that person can expect that knowledgeable FITS people will be listening. So, if you have something you want to say, go ahead.*
*– Don Wells [27]*

By 1991, the pressing issues FITS addressed in the 1970s – file interchange, in particular – acquired new answers as computing was changed by the network structure of the internet. The computing ground had started to shift as the problem FITS was designed to solve faded into the past, and new problems, and potential new answers, came to the fore. The internet itself would soon evolve from a space of occasionally comical and often frustrating misunderstandings into a fully formed digital infrastructure capable of coordinating far flung collaborators and supplementing the scholarly societies and institutions that had sped FITS diffusion through astronomy in the 1980s.

### 3.1 Here comes _everyone_,_everywhere_

As the internet gained wider adoption in the late 1980s and early 1990s, the impetus to share progress and thoughts about FITS on the internet grew. There were many options. Usenet news groups were a public way to post and discuss news that anyone could join. Listservs were a less public method of sharing information; via an email server, pre-approved listserv members could create posts and exchange information via threads. Augmenting listservs were email exploders, programs that allowed an email sent to a particular address to be forwarded or exploded to multiple email addresses. Wells started with a news group and eventually linked it via email exploder to a listserv, pushing FITS communications to both the public via newsgroup and to listserv members via email. For sake of brevity, and the fact that the content of the newsgroup, associated listservs and the fitsbits email exploder were closely interrelated, we will take liberties and call the intertwined messaging systems a listserv.

Listservs became popular with scientific groups in the early 1990s, as both an informal and formal medium, a way to banter and to coordinate formal actions. For FITS, the listserv bridged the space between hallway conversations in academic departments and observatories and official communications of the IAU and ADASS. The listserv was one of a bundle of FITS services introduced in the late 1980s and early 1990s designed to increase the public presence of FITS on the internet: online software directory, ftp archive, and e-mail exploder.

On May 30th 1991, Don Wells sent the initial message to the FITS usenet group.

```
Newsgroups: alt.sci.astro.fits
Distribution: alt
Organization: National Radio Astronomy Observatory, Charlottesville, VA
From: dwells at fits.cx.nrao.edu (Don Wells)
Subject: Activation of alt.sci.astro.fits
Date: Thu, 30 May 1991 06:27:50 GMT

I have formed this group in order to provide a forum for discussion of
technical issues related to the FITS [Flexible Image Transport System]
data format, the standard data interchange format of astronomy
worldwide.  The idea of a newsgroup for FITS was recommended by Trond
Melen of the European Southern Observatory in Garching, FRG.

I discussed the concept of this newsgroup in my annual presentation on
the state of FITS, made in the morning business session of the Working
Group for Astronomical Software held on Tuesday 28th May (yesterday),
at the summer meeting of the American Astronomical Society in Seattle.
I did not sense any opposition to the idea among the people present at
the WGAS meeting, and so the group has been created.

I do not intend to make any more transmissions on this newsgroup until
Preben Grosbol (Chairman of the IAU [International Astronomical Union]
FITS Working Group) has posted to the newsgroup. I would also like to
see a posting by Barry Schlesinger (the FITS "hotline" person at the
FITS Office at NSSDC-NOST).
```

Fig. 2. A snippet of the first alt.sci.astro.fits message [27]

In a continuation of the collegial academic style of peer committees and BoF groups, Wells demurred discussing FITS further until his colleagues Preben Grøsbol and Barry Schlesinger had weighed in on usenet. But the internet waits for nobody and it wasn't long before "_everyone_everywhere_ " joined the news group. In contrast to formal communication style of academic governance, the news group sometimes brought hot "flame wars" into the discussion over FITS' future [28].

```
>I'm in the Space Shuttle Program Office (Level II), in the
>Management Integration/Information Systems Office at JSC.  My
>office sets information systems standards for the Space Shuttle
>Program.  My boss asked me to put together a draft of a photo
>transmission and storage standard.
>
>I proposed to the Internet that I was going to make the Shuttle program
>standardize on GIF as the image file storage and interchange format.
>Everybody said no, so I changed by mind and proposed TIFF 5.0.  The
>plan is to impose an interim standard now, then move all our files to
>the official US Gov't standard when it comes out.

I don't flame in public that often, but this is above and beyond....

I didn't reply to your original post because *I THOUGHT IT WAS A JOKE*!

I simply couldn't believe that a) someone would possibly suggest using
GIF as a serious format for interchange of even semi-scientific data
and that b) someone would, in the same message, suggest that this
decision would have serious implications across the entire US space
program, ask questions of a level indicating that essentially *no*
background research has been done ("What are the advantages and
disadvantages of GIF vs. the other color image standards?") and
propose as the standard the single format about which he had
(incomplete) information.
```

Fig. 3. A bit of ribbing on the listserv [29]

In addition to being a departure from the formal style of academic governance, the comments underline the kind of expertise and tacit knowledge [30] assumed to be shared within the confines of academic governance. Anyone with even a passing familiarity with astronomical imaging would know intuitively why GIF and TIFF are inadequate for transmitting astronomical data. Hence, an unintended consequence of the news group was to open a window for all manner of comments, many of which had to be addressed and rebutted.

Despite the novel inquiry about GIFs and TIFFs addressed to "the internet", the listserv was comparatively free of the "flame wars" that plagued many listservs and usenet groups. Most of the disagreements were about technical specifications, such as which keyword headers should be standardized or suggestions for implementing a coordinate system within FITS. As we discuss in the following section, these questions would prove to be more consequential than inquiries about GIFs and TIFFs.

## 4 Problems locating the World Coordinate System

*"Good form in FITS always extends to include the human as well as the software readers". [31]*

By 1998, FITS was well-entrenched as the default file format of astronomy. The email exploder was working as Wells intended, with various discussions and threads explaining what FITS is, how it can be used, and proposed changes in the future BoF sessions. But in 1998, a lingering and unresolved topic necessitated action: The World Coordinate System (henceforth WCS). In the WCS controversy, the semantic elements necessary to interpreting data began to cut against FITS' syntactic function as an interchange format.

World Coordinates serve to situate an object in a space, so that its location can be fixed. The WCS is a set of transformations that map locations in the sky to pixel locations in an image, or the reverse, from pixel images to positions in the sphere of the sky. These locations, or measurements at locations, can be multi-dimensional in form [32]. The WCS can also be used to define wavelength transformations, for example for spectroscopy, the study of electromagnetic radiation emitting from stars.

The original FITS specifications from 1981 described a method for locating the coordinates of image pixels but did not specify conventions for how to locate those coordinates in the sphere of the sky. This was a deliberate design decision that left the astronomer to decide how to map the image to the sky. Over time, differing groups of astronomers (optical, radio, X-ray, infrared, ground-based and space-based) had developed de facto, and incompatible, conventions for representing the sphere of the sky, cutting against FITS mandate as an interchange format. To solve some of these issues caused by the lack of definition in the FITS standards, some conventions were created by members of the FITS community that were non-standards yet widely adopted, such as ad hoc versions of the WCS. Software libraries popped up to help with these coordinate transformations needed to interpret FITS images and fill in gaps left by the official FITS standard [33].

Wells' solution to the WCS problem was to create a listserv called wcsfits, to build consensus for a standardized approach in FITS using the WCS by including input from all observing astronomers in the proposals for addendums to FITS. The wcsfits listserv revealed a chaos of competing opinions about how a standardized WCS might be implemented. Implementing any coordinate system is not a trivial undertaking as the difficulty lies in the nature of locating an image in the sky, not in representing it in the format. Vexing issues arose as to how the tools could read the WCS components in the FITS files: should FITS ignore some methods and read others? How much complexity should be built into FITS headers? The disagreements lie in both the organizational structure of the file and the differing ways groups of astronomers had organized their measurements and orientations towards the sky. Throughout the late 1990s, various approaches were discussed at length on the wcsfits listserv, without a clear consensus.

Yet by 2002, an initial attempt at a standard approach had been made. Two papers were published in 2002, one in 2006, and another in 2015 [34]. These new WCS conventions provided a standard method to map physical coordinates in the sky. Though the WCS issues played out for years (and still do) within FITS governance, and certain functionality has been added, for most it still requires effort on the part of the researcher to make it work.

Griesen [35] argues that the failure to adopt a WCS system in FITS is due to lacking the will to settle on something. FITS critics seem to think that the failure to adopt WCS guidelines is another argument against FITS. WCS exposes the differences in subfield work and pushes FITS to the existential limit of what FITS should *be*- should FITS be a Swiss army knife or provide directions to build one? It was too late, though, unlike the way FITS was developed (by a handful of people), using the listserv to create consensus for a standardized way to represent the WCS was a failure. By the time a standard was proposed, subfield-specific methods of implementing the WCS were entrenched and many in astronomy had moved on to new forms of astronomical analysis.

## 5 Analytic pipelines and the return of the interchange problem

*"Once FITS, always FITS" doctrine, which has been utilized to effectively freeze the format, was a mistake in our opinion. Adherence to the doctrine, and lack of any means to version the format in a machine-readable manner, has stifled necessary change of FITS. [36]*

As FITS transitioned from an interchange format capable of bridging subfields and computing systems into the de facto archiving standard for astronomical data, new astronomical tools, such as Infrared Reduction and Analysis Facility (henceforth IRAF) [37] were built on top of FITS. Meanwhile, tools imported from outside astronomy, such as relational databases that served FITS files, and object-oriented programming languages that took FITS as objects, became commonplace within astronomy. These changes were driven by what has come to be called data-intensive science, which has seen the velocity and volume of data increase by several orders of magnitude since 1981 when astronomers went up the mountain to observe and came down the mountain with a box full of magnetic tapes to analyze. Today, astronomers are more likely to divert a stream of data from a server. Consequently, many astronomical research groups now employ statisticians, data scientists, and archivists simply to stay afloat on the ever-increasing flow of data in astronomy [38]. And rather than being a question of whether or not the magnetic tapes one brings off the mountain are compatible with one's institution's computers, analysis takes the form of complex pipelines in which data is transformed and reduced in a series of steps.

The WCS controversy illustrated why FITS is not an ideal format for analytical work in astronomy today. Several recent papers have outlined issues in current computational environments in which FITS does not perform well [36]. For example, if one has a FITS file of 100GB, which is larger than the RAM memory in the computer, the file will be read slowly. To quickly read large files, a modern container is required to handle a large array of data. Astronomers introducing new kinds of analysis wanted more analytical options [39], in particular, they chafed against the FITS structure that determines when analysis takes place [35]. As another astronomer noted, FITS works elegantly with the volume and velocity of data common in the 1980s and 1990s, but warned that "as data sets get bigger it is going to start failing." [40].

Data sets have gotten bigger and the increasing volume and velocity of data within astronomy has seen the interchange problem return in a surprising way – as a new chaos of programming languages and file formats deployed in astronomical data pipelines. A data pipeline is a set of actions to designed to extract and transform data for analytical use [41]. The interchange problem increases as data moves further down the pipeline, where the the needs and inclinations unique to a research group accumulate in the bits and pieces of programming languages and formats a pipeline is constructed from. FITS is still omnipresent, but today often serves as a point of departure, not an endpoint.

Three file formats in particular, each using FITS as a touchstone, are in use within astronomy: the Hierarchical Data Format (henceforth HDF5) stemming from federally funded organization attempts at creating a universal data analysis format, VOTable, an attempt at building a broad set of XML-based tools for astronomical archiving and web-based analysis, and Advanced Scientific Data Format (henceforth ASDF) a newer file format with many similarities to FITS, such as a human readable headers with binary structures for data. HDF5 is used throughout many scientific fields for high speed computing. HDF5 emerged out of a score of file formats proposed for use in the Earth Observing System project headed by NASA [42] and over time matured into a format able to store different kinds of information. It can handle large files in the terabyte range and is therefore favored by many data scientists. VOTable format was designed for astronomers to continue the work of the Virtual Observatory methods of making online data interoperable [43]. VOTable uses XML as a standard to represent data as a set of tables, with XML being the interoperability piece. However, it has a major drawback, in that VOTable has no method to handle binary data. While HDF5 addressed the needs of data scientists and VOTable the needs of taking astronomy online for engaging citizens in astronomical analysis [44], ASDF was developed as a direct replacement for FITS.

Publically announced in a 2015 paper [45], like FITS, ASDF was created by astronomers [46]. ASDF resembles FITS in structure and function, containing a human and machine-readable header, rendered in YAML and is backwards compatible with itself for archival purposes. Unlike the FORTRAN-derived FITS headers, however, YAML headers support hierarchical information, thus overcoming FITS' limitations in working with WCS coordinates.

Unlike during the 1970s, today computers are highly interoperable. Yet within astronomical pipelines, software interdependencies, software libraries relying on other software libraries for basic functions, have made for extremely complex layers of code that are brittle and prone to breakage [47]. One widely used strategy to overcome this problem is organizing software infrastructure via an open source Github repository, where interdependency problems can be crowdsourced. Like FITS, ASDF is reliant on a tremendous amount of volunteer work. On the ASDF Github repository, volunteers are welcomed in name of openness in astronomy and encouraged to make contributions. The Github review mechanism sets up a system of governance, where contributors to the project are welcomed, but their contributions might be rejected, ignored, or folded into the project by a closed circle of core developers. In a sign of emerging institutional acceptance, NASA's delayed but hopefully soon to be launched James Webb Space Telescope will offer data in both ASDF and FITS formats. Since the focus of money spent on astronomy is geared towards the instrumentation, this institutional support augers well for ASDF's long-term future, and points towards the possibility of institutionally based governance in ASDF's future.

## 6 THE FUTURE(S) OF ASTRONOMICAL INFRASTRUCTURE?

*Code is hard, code rots. So if your data depends on having code available like this opaque binary file, that's bad. And that's where astronomy... It's FITS files. Well, not everything needs to be in a FITS file. It certainly lets you package up both data so that they will be around for a long time. And I hear this in other fields like "we don't have anything like that, how do you guys do it?" Physics, we don't have anything like that. Biology, we don't have anything like that. It's all in people's notebooks whatever. – research scientist at a U.S. research lab [48]*

Created by and for astronomical observers in the late 1970s as a means to share files across notoriously non-interoperable computers, FITS is a simple format consisting of an ASCII header containing metadata, along with optional extra units [49], and data units consisting of images and tables of ASCII or binary data. FITS has managed to serve as an interchange format between machines, subfields, and international differences. Though informal and rickety (agonizingly slow for some, overly stable for others), FITS remains the de facto standard for archival and interchange in astronomy today [50]. It is estimated that there are more than 1 billion FITS files in various ar-chives around the world [39]. Like the format itself, which is flexible enough to accommodate radio astronomers, optical astronomers, and Vatican archivists, and durable enough to ensure the original FITS files produced forty years ago are still usable, the infrastructure of FITS governance has proven flexible and durable enough to see FITS through sea changes in the nature of computation and astronomical research.

Astronomers who venture into other fields often find themselves mired in a chaos of competing proprietary formats that mirrors the situation in astronomy during the 1970s. As one astronomer replied when asked what astronomy would be without FITS, "it would be like materials science is today… if there are 10 different vendors' machines down the hallway here, likely 10 different formats are coming out of them, not interoperable, not with common metadata standards" [51].

FITS began its life as an interchange and subsequently an archive format, but inevitably it also became a format for analysis. As a format for analysis, FITS is tied to a particular computing paradigm, imperative programming, and the computer language, FORTRAN, that were the common currency of astronomy in the 1970s. It has become antiquated by $21^{st}$ century standards, reflecting the technical norms of the late 1970s, and is quickly being supplanted by newer paradigms and languages. Further, FITS is also tied to a particular governance infrastructure that was common currency in the academia of the late 1970s, but seems increasingly antiquated given changes to both the discipline of astronomy and working conditions in academia. Yet paradoxically, in the era of data-intensive science, FITS has become increasingly useful to astronomy as researchers mine archives for datasets from different instruments and missions and combine them in new ways to produce new research.

What does it mean for FITS governance that FITS is both increasingly useful and increasingly antiquated? The process for amending and adding to the standard assures broad community participation, and although this sometimes makes the process of change rather slow it helps to assure community support and compliance [52]. So, FITS carefully crawls along as it has for decades, the working group is scheduled to meet at ADASS in October 2018 to look over proposed modifications to FITS.

Infrastructure for scientific disciplines that is expected to span generations requires care and attention to both the technical details and governance infrastructure. It must be attentive to, and change, with the context of the academy and specific disciplines. Can a cohort similar to the FITS cohort be formed, and if so, what form would it take? Is it possible to replicate the kind of stability long academic careers and stable academic associations gave to FITS through ad hoc projects organized over the internet? No single organization is in charge of thinking through next generation astronomy standards. Today, the precarious postdoc and research scientists who do the everyday work of scientific computation lack the time, and increasingly the incentive, to volunteer for academic governance work like improving and maintaining FITS, in a profession they may leave after a few years. The biggest threat to FITS governance and, therefore, its future in astronomy, might be the upcoming generational turnover in the committees and working groups that care for, maintain, and update FITS. Despite the uncertainty of FITS future care and governance and the emergence of new forms of astronomical analysis, the observer's wisdom embodied in the edict "once FITS, always FITS" remains as relevant as ever to astronomy.

Perhaps it is not surprising, then, that the newest file format in astronomical computing today takes inspiration from the oldest. ASDF, created by astronomers for astronomy, can be understood as a technical update of FITS set on a new governance infrastructure with a new collaboration and governance style more in keeping with the norms of open source software communities. Despite ASDF's intention to be a modern update of FITS, it shouldn't be viewed as a replacement for FITS. Millions of FITS files are stored on servers across the world. They are openable, viewable, usable, to the degree that their creators specified the metadata in the header in a way that can be understood by others. As astronomy archives' usage increases, as evidenced by citations in publications, FITS files themselves are increasingly useful to astronomers.

## ACKNOWLEDGMENT

This research is funded by the Alfred P. Sloan Foundation, Award# 2015-14001: If Data Sharing is the answer, what is the question? We thank Christine Borgman, Milena Golshan, Morgan Wofford, Peter Darch, and Cheryl Thompson of the UCLA Center for Knowledge Infrastructures. We also thank and Tuan Do, Mark Morris, Abhimat Gautam and all members of the Galactic Center Group at UCLA for additional astronomy research.

**Michael Scroggins** graduated from Columbia University with a PhD in Cultural Anthropology. He is currently a postdoctoral scholar at the Center for Knowledge Infrastructures, University of California Los Angeles.

**Bernadette M. Boscoe** is a PhD candidate in Information Studies at UCLA. She is currently a member of the Center for Knowledge Infrastructures at UCLA. Her current research concerns keeping scientific data alive over decades.